\documentclass[pre,amsmath,amssymb,twocolumn,nofootinbib]{revtex4-1}
\usepackage{graphicx}
\usepackage{bm,color}
\usepackage{times}
\usepackage[colorlinks=true,linkcolor=blue,urlcolor=blue,citecolor=blue]{hyperref}

\newcommand{\ca}[1]{{\cal #1}}
\newcommand{\eq}[1]{(\ref{#1})}
\newcommand{\Eq}[1]{Eq.~(\ref{#1})}
\newcommand{\Eqs}[1]{Eqs.~(\ref{#1})}
\newcommand{\nn}{\nonumber}

\newcommand{\rmd}{\mathrm{d}}
\newcommand{\rme}{e}

\newcommand{\be}{\begin{equation}}
\newcommand{\ee}{\end{equation}}
\newcommand{\beq}{\begin{equation}}
\newcommand{\eeq}{\end{equation}}
\newcommand{\bea}{\begin{eqnarray}}
\newcommand{\eea}{\end{eqnarray}}

\usepackage{graphicx}

\newcommand{\fig}[2]{\includegraphics[width=#1]{#2}}

\newcommand{\Fig}[1]{\includegraphics[width=\columnwidth]{#1}}

\newcommand{\checked}{}

\begin{document}

\title{Span observables --   ``When is a foraging rabbit no longer hungry?''}

\author{Kay J\"org Wiese}
 \affiliation{Laboratoire de Physique de l'Ecole normale sup\'erieure, ENS, Universit\'e PSL, CNRS, Sorbonne Universit\'e, Universit\'e Paris-Diderot, Sorbonne Paris Cit\'e, Paris, France.}

\begin{abstract}
Be $X_t$ a random walk. We study its span $S$, i.e.\ the size of the domain visited up to time $t$.  We want to know the probability that $S$   reaches $1$ for the first time, as well as the density of the span given $t$. 
Analytical results  are presented, and checked against numerical simulations. We then generalize this to include drift, and one or two reflecting boundaries.  We also derive the joint probability of the maximum and minimum of a process. Our results are based on the diffusion propagator with reflecting or absorbing boundaries, for which a set of useful  formulas is derived. 
\end{abstract}
\maketitle

\section{Introduction}\begin{figure}[b]
\includegraphics[width=\columnwidth]{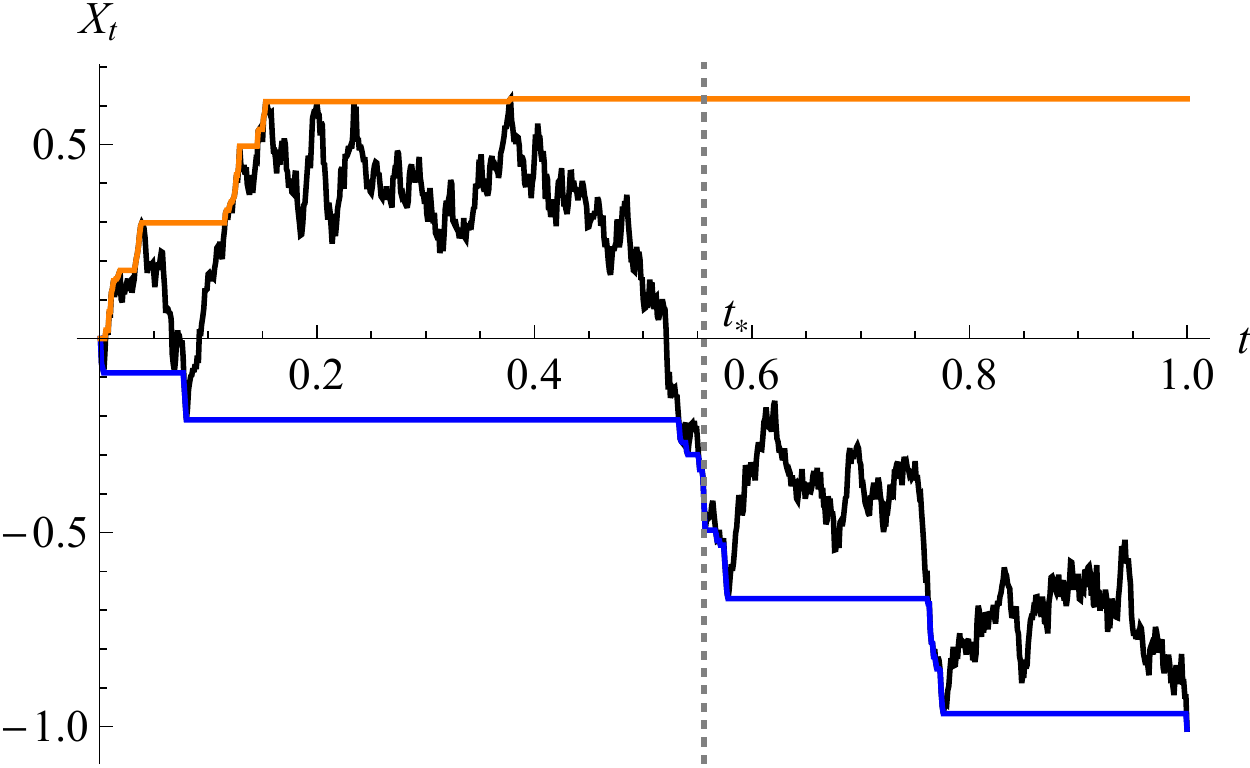}
\caption{The random process $X_t$, with its running max (in orange) and min (in blue).} 
\label{f:running-max-and-min}
\end{figure}
Consider a Brownian motion $X_t$,  starting at $X_0=0$, with drift $\mu$, and variance 2, 
\begin{align}
&\left< X_t\right> = \mu t\ . \\
\label{variance}
& \left< \big(X_t-\left< X_t\right>\!\big)^{\!2}\right> = 2 t\ .
\end{align}
A sample trajectory is sketched on Fig.~\ref{f:running-max-and-min} (for $\mu=0$).
A key problem in  stochastic processes are  the first-passage properties \cite{FellerBook,RednerBook} in a finite domain,  say the unit interval $[0,1]$.
For a Brownian,  the probability to exit at the upper boundary $x=1$ without   visiting the lower boundary at $x=0$, while starting at $x$ is  
\be\label{1}
P_{1}(x) = x\ .
\ee
Another key observable is the exit time, starting at $x$, which 
behaves as   $\left< T_{\rm exit}(x)\right>_0 \sim x(1-x)$. 

Here we consider a different set of observables, namely the {\em span} of a process: Define 
 the positive and negative records (a.k.a.\ the running max and min) as 
\bea
M_+(t) &:=& \max_{t'\le t}  X_{t'} \ ,\\
M_-(t) &:=& \min_{t'\le t}  X_{t'} \ .
\eea
These observables are  drawn on Fig.~\ref{f:running-max-and-min}. 
The span $S(t)$ is their difference, i.e.\ the size of the (compact) domain visited up to time $t$, 
\be
S(t):= M_+(t) - M_-(t)\ .
\ee
We   study  the probability that $S(t)$ becomes 1 for the first time. Curiously, this observable is rarely treated in the literature, and most of the studies we found  are   concerned with questions of convergence of the first moments, which is non-trivial when the process is more complicated than a random walk: Let us mention the mean first-passage time \cite{WeissDiMarzioGaylord1986}, with some discrepancies stated in Ref.~\cite{PalleschiTorquati1989}. The full distribution as a function of times is derived below.  
A related  but distinct observable is the density of the span at time $t$, considered in the classic references \cite{Daniels1941,Feller1951,WeissRubin1976,PalleschiTorquati1989}.
A beautiful recent result is the covariance of the span \cite{AnnesiMarinariOshanin2019}.

\begin{figure}[b]
\Fig{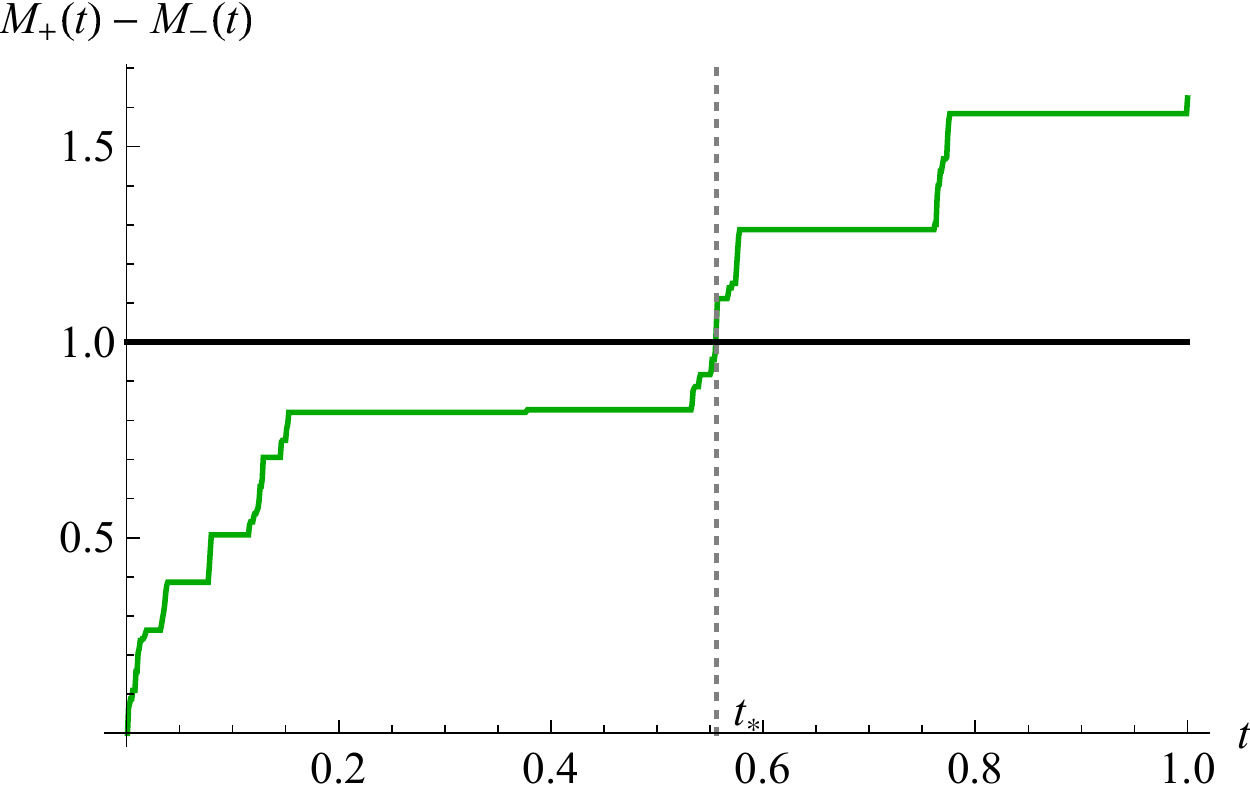}
\caption{The span $S(t):=M_+(t)-M_-(t)$ of  $X_t$ from Fig.~\ref{f:running-max-and-min}.}
\label{f:span}
\end{figure}

One may ask where span observables actually occur in nature? One example is the {\em Hungry Rabbit Problem}.  Suppose a hungry and myopic rabbit is released. It will perform a random walk, until its stomach is full, i.e.\ the span of its trajectory reaches $1$. This is a variant of the myopic rabbit introduced in \cite{RagerBhatBenichouRedner2018}. We will give the probability for the time that the rabbit is no longer hungry analytically, including some drift in the rabbit's motion, when e.g.\ it prefers to move downhill. 
One may object that the problem is not realistic, since foraging the rabbit consumes food. 
We currently have no solution for the latter problem, even though
diffusion with moving boundaries can, at least in principle, be treated via a set of integral equations \cite{Cannon1984}; however we do not know of a closed-form solution. A notable exception are expanding boundaries in  the limit of  large times, where the survival probability can be evaluated analytically 
\cite{BraySmith2007}.

Another example  arises in measuring the exit probability from a strip, a problem studied for fractional Brownian motion in Ref.~\cite{Wiese2018}. The question is how long one has to run a simulation until the process $Y_t:=X_t+x$ has exited from the unit interval $[0,1]$, for all $x\in [0,1]$. We claim that the simulation can   be stopped at time $t_*$ when the span first reaches 1. To understand this statement, define $x^*:=-M_-(t^*)$. Then consider the process $Y_t:=X_t+x$. If $x>x^*$, then $Y_t$ will exit at the upper boundary at time $t< t^{*}$, while for $x<x^*$ the process $Y_t$ will exit at the lower boundary at time $t< t^{*}$. For $t<t_*$ there will be  some $x\approx x_{*}$ where no conclusion can be drawn. Thus $x_*$ samples $P_1'(x)$, the derivative of the exit probability at the upper boundary,  and $t_*$ is the time one can stop the process without loss of information.

A related quantity is the joint density of the running maximum and minimum, given $t$. This question is relevant in the analysis of stock-market data \cite{WergenBognerKrug2011}, where it allows one to quantify violations of the Markov property. 

The span  is also relevant in the search of a protein for its binding site on a DNA molecule. The idea of {\em facilitated diffusion} \cite{MirnySlutskyWunderlichTafviziLeithKosmrlj2009} is to alternatively diffuse along the DNA molecule or in 3d space, thus optimizing the search.

Finally, the span is not a Markov process, but a process with memory, as it remembers its positive and negative records. This places our study in the larger context of processes with memory, of which fractional Brownian motion may be the most relevant one \cite{NourdinBook,DiekerPhD,Krug1998,SadhuDelormeWiese2017}. 
 
 The remainder of this article is organized as follows: We first derive key results for Brownian motion in the unit interval $[0,1]$, with absorbing boundary conditions at both ends, see section \ref{s:Basic formulas for Brownian Motion with two absorbing boundaries}. This is then generalized to one absorbing and one reflecting boundary in section \ref{s:prop-one-ab+one-reflecting}, and to two reflecting ones in section \ref{s:prop-two-reflecting}. Most of these results are  known. 
We   give analytical results for span observables in section \ref{s:Probabilities-for-the-span}, and the joint distribution of running maximum and minimum in section \ref{s:joint-max+min}.
A generalization to a random walk with   one reflecting boundary is presented in section \ref{s:Diffusion with a reflecting wall}, while two reflecting boundaries are treated in section \ref{s:The span with two reflecting boundaries}.
We conclude in section \ref{s:Conclusions and Open Problems} with   open problems. 

\section{Basic formulas for Brownian Motion with two absorbing boundaries}
\label{s:Basic formulas for Brownian Motion with two absorbing boundaries}

\subsection{Solving the Fokker-Planck equation}
Consider a random walk $X(t)$ given by its Langevin equation 
\be
\partial_t X(t) = \mu t +  \eta(t)\ , \qquad \left< \eta(t ) \eta(t')\right> =2   \delta (t-t')  \ .
\ee
There are absorbing (Dirichlet) boundary conditions both at $x=0$, and $x=1$. 
If the trajectory starts at $X_0=x$, and ends at $X_t=y$, 
then the forward Fokker-Planck equation reads \cite{FellerBook,RednerBook}
\be\label{forwardFP}
\partial_t P_{\rm DD}^\mu(x,y,t ) =  \frac{\partial ^2}{\partial y^2}  P_{\rm DD}^\mu(x,y,t) - \mu  \frac{\partial }{\partial y}  P_{\rm DD}^\mu(x,y,t) \ .
\ee
The index ``DD'' refers to the two absorbing (Dirichlet) boundary conditions at $x=0$, and $x=1$. The probability to survive at time $t$  is given by $\int_0^1\rmd y\, P_{\rm DD}(x,y,t)$.
The  general solution of the Fokker-Planck equation \eq{forwardFP} can be written as
\be\label{P+mu(x,y,t)}
{P_{\rm DD}^\mu(x,y,t)} =\rme^{\frac{\mu (y-x)}2 -\frac{\mu^2 t}{4}}  \left[  {\mathbb P}(x-y,t)- {\mathbb P}(x+y,t) \right]\ .
\ee
The key object in this construction is  
\be\label{mathbbP}
{\mathbb P}(z,t):=\frac{1}{\sqrt{4\pi t}} \sum_{n=-\infty}^{\infty} e^{-(z+2n)^2/4t} = \frac{1}{2} \vartheta _3\!\left(\frac{\pi}{2}   z,e^{-\pi ^2 t}\right).
\ee
$\vartheta$ is the elliptic $\vartheta$-function. 
Using the Poisson summation formula, an alternative form for  ${\mathbb P}(z,t)$ is
\be\label{PP-Poisson}
{\mathbb P}(z,t) = \frac12 + \sum_{m=1}^\infty \rme^{-m^2 \pi^2 t} \cos(m \pi z)\ .
\ee
To prove the above  statements it is enough to remark that \Eq{P+mu(x,y,t)} satisfies the Fokker-Planck equation (\ref{forwardFP}),  vanishes at $y=0$ and $y=1$, and reduces for $t\to 0$ to a $\delta$-function
\begin{equation}\label{f1}
\lim_{t\to 0}P_{+}^\mu (x,y,t) = \delta (x-y)\ .
\end{equation}
The function $\mathbb P(z,t)$ has the following properties
\be
\mathbb P(z,t) = \mathbb P(z+2,t) = \mathbb P(-z,t) \ .
\ee
As a consequence, 
\be
\partial_z \mathbb P(z,t) |_{z=\rm integer}= 0
\ee
It is useful to   consider its Laplace-transformed version. 
We define the Laplace transform of a function $F (t)$, with
$t\ge 0$, and marked with a tilde  as 
\begin{equation}\label{f56-bis} 
\tilde F (s):= \mathcal{L}_{t\to s } \left[ F (t) \right]  
=  
\int_{0}^{\infty}\rmd t\,  e^{-s t} F (t) \ .
\end{equation}
This yields for $-2<z<2$
\bea
\tilde {\mathbb P}(z,s) &=& \frac{e^{-\sqrt{s} |z|} }{2
   \sqrt{s}} + \frac{\left[\coth (\sqrt{s})-1\right] \cosh
   (\sqrt{s} z)}{2 \sqrt{s}} \\
   &=& \frac{1}{2 s}+\frac{1}{12}  (2-6 \left| z\right| +3
   z^2 ) \\
   && +\frac{s}{720}  \left(-60 z^2 \left| z\right| +15
   \left(z^2+4\right) z^2-8\right)+...\nn
\eea
And
\be
\mathcal{L}_{t\to s } \left[ \rme^{ -\frac{\mu^2 t}{4}}    {\mathbb P}(z,t) \right] = \tilde {\mathbb P}\Big(z,s+\frac{\mu^2}4\Big)\ .
\ee
Note that the combination in square brackets in \Eq{P+mu(x,y,t)} can also be written as 
\begin{eqnarray}\label{16-bis}
\lefteqn{ {\mathbb P}(x-y,t)- {\mathbb P}(x+y,t)}\nn\\  &=& \frac{e^{-\sqrt{s} |x-y|}-e^{-\sqrt{s} (x+y)}}{2
   \sqrt{s}} \nn\\
   && -\frac{\left[\coth  (\sqrt{s})-1\right] \sinh
   (\sqrt{s} x) \sinh (\sqrt{s} y)}{\sqrt{s}}.~~~~~~
\end{eqnarray}
The form \eq{16-bis}   facilitates its integration over  $x$ and $y$, which is useful when concatenating several propagators \cite{Wiese2018}.

\subsection{Boundary currents and conservation of probability}

Conservation of probability reads (the variable $x$ is   the initial condition, here a dummy variable)
\be\label{current-conservation}
\partial_t P_{\rm DD}^\mu(x,y,t) +\partial_y J_{\rm DD}^\mu(x,y,t) = 0\ .
\ee
$J^\mu_{\rm DD} $ is the current, which from \Eqs{forwardFP}, \eq{P+mu(x,y,t)} and \eq{current-conservation} can be identified as 
\bea\label{22}
&&J_{\rm DD}(x,y,t)= ( \mu-\partial_y) P_{\rm DD}^\mu(x,y,t) \\
&&\qquad = \rme^{\frac{\mu (y-x)}2 -\frac{\mu^2 t}{4}} \Big(\frac\mu2-\partial_y\Big) \left[  {\mathbb P}(x-y,t)- {\mathbb P}(x+y,t) \right]\ . \nn
\eea
Due to the Dirichlet conditions at $y=0$ and $y=1$, we have 
\be
\int_0^1 \rmd y \, \partial_t P_{\rm DD}^\mu(x,y,t) = J_{\rm DD}^\mu(x,0,t) - J_{\rm DD}^\mu(x,1,t)\ .
\ee
Thus,  
the probability to exit at time $t$, when starting in $x$ at time $0$ reads
\begin{align}
& P_{\rm exit}^{\rm DD}(x,t)=  J_{\rm DD}^\mu(x,1,t) - J_{\rm DD}^\mu(x,0,t) \nn\\
&= 2\,\rme^{-\frac{\mu ^2 t}{4}} \left[   e^{\frac{\mu (1-x)}{2}}\partial_x \mathbb P(1-x,t) -e^{-\frac{\mu  x}{2}}\partial_x \mathbb P(x,t) \right] .
\end{align}
The outgoing currents at the upper and lower boundary are 
\bea
J_{\rm DD}^\mu(x,1,t)  &=&   2\,\rme^{-\frac{\mu ^2 t}{4}}     e^{\frac{\mu (1-x)}{2}}\partial_x \mathbb P(1-x,t)\ ,  \\
-J_{\rm DD}^\mu(x,0,t) &=& -  2\,\rme^{-\frac{\mu ^2 t}{4}}    e^{-\frac{\mu  x}{2}}\partial_x \mathbb P(x,t) \ .
\eea
The Laplace transforms  of these outgoing currents are 
\bea
- \tilde J_{\rm DD}^\mu(x,0,s) &=& e^{-\frac{\mu  x}{2}} \frac{ \sinh \big(\sqrt {s'}  (1-x)\big) }{\sinh(\sqrt {s'} )}\Bigg|_{s'=s+\mu^2/4}\ , \qquad \\
\tilde J_{\rm DD}^\mu(x,1,s) &=& e^{\frac{\mu (1- x)}{2}}  \frac{ \sinh (\sqrt {s'}\, x) }{\sinh(\sqrt {s'} )} \Bigg|_{s'=s+\mu^2/4}\ .
\eea

\begin{figure*}[t]
\fig{8.1cm}{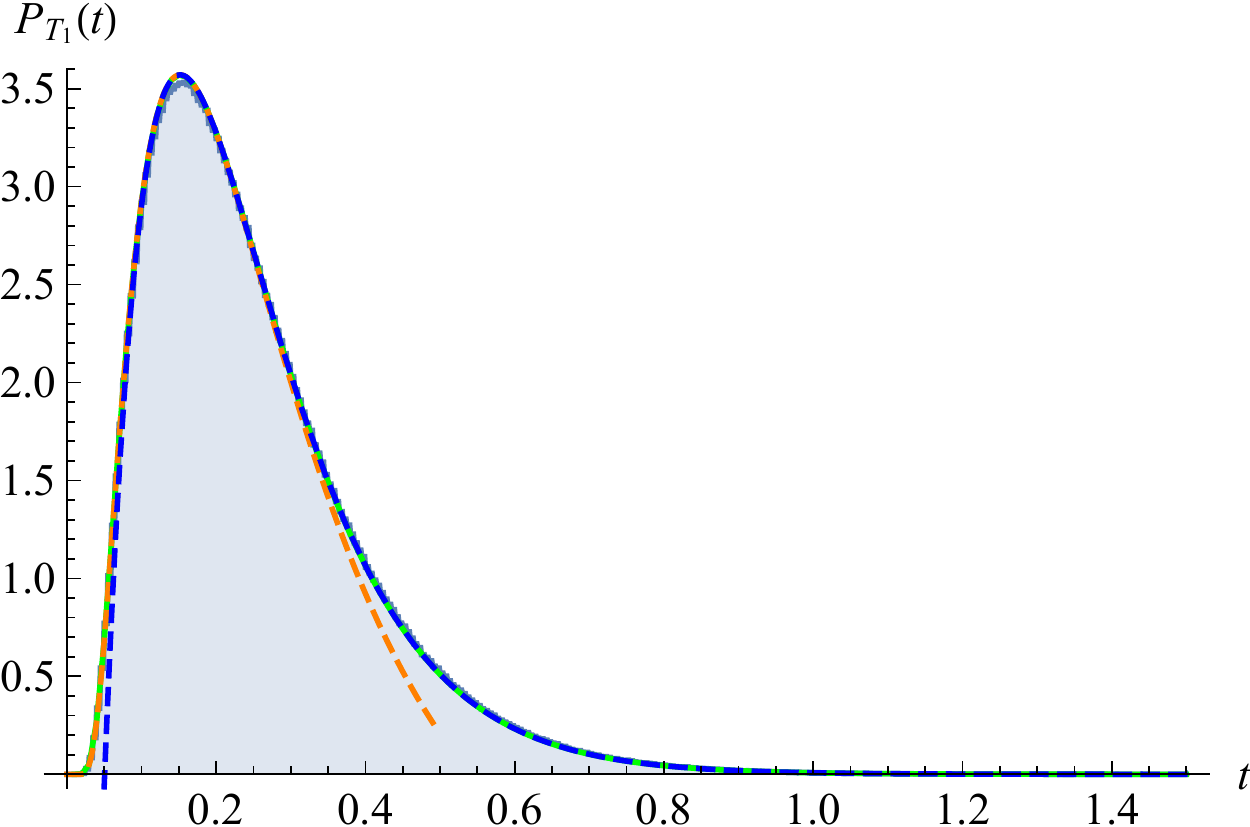}~~~~~~~~~\fig{8.6cm}{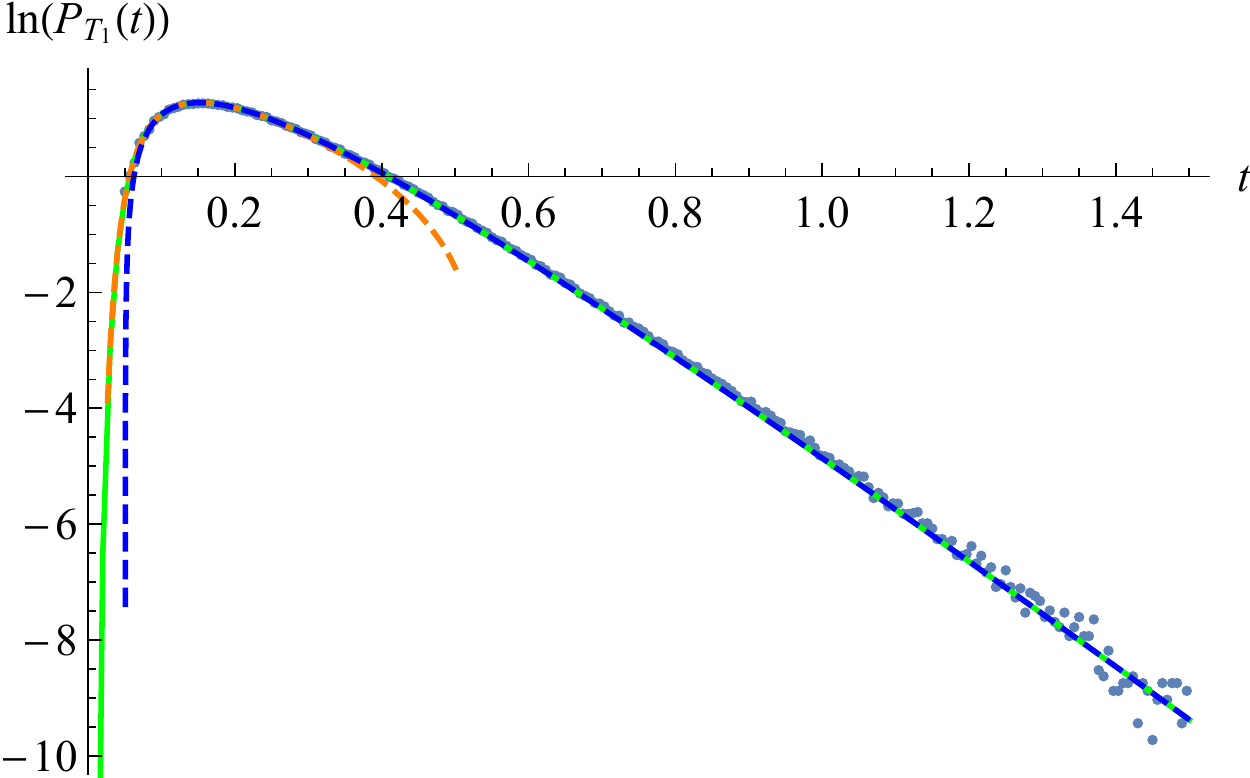}
\caption{The probability that the span reaches 1 for the first time. Grey: RW simulation with $\delta t=10^{-5}$, and $10^6$ samples. Green: the analytic result (\ref{PT1}). 
Orange dashed the small-times asymptotics (\ref{PT1-small}); blue dashed the large-time asymptotics (\ref{PT1-asymp}). Note a small systematic deviation due to the relatively large time step $\delta t=10^{-5}$. }
\label{f:Pspan}
\end{figure*}

\subsection{Absorption probabilities at $x=0$ and $x=1$}
The absorption probabilities at $x=0$ and $x=1$ are  
\bea\label{14}
P_{\rm DD,0}^\mu(x) &:=& \int _0^\infty \rmd t \, \left[ -  J_{\rm DD}^\mu(x,0,t) \right] = \lim_{s\to 0}  \left[ - \tilde J_{\rm DD}^\mu(x,0,s) \right]\nn\\
& =&   e^{-\frac{\mu  x}{2}}
   \frac{\sinh \left(\frac{\mu}{2}   (1-x)\right)}{\text{sinh}\left(\frac{\mu }{2}\right)} \ , \\
\label{15}
P_{{\rm DD},1}^\mu(x) &:=& \int _0^\infty \rmd t \,    J_{\rm DD}^\mu(x,1,t) = 
\lim_{s\to 0}    \tilde J_{\rm DD}^\mu(x,1,s) \nn\\
&=&  e^{-\frac{1}{2} \mu 
   (x-1)} \frac{\sinh \left(\frac{\mu  x}{2}\right) }{\text{sinh}\left(\frac{\mu }{2}\right)}\ .
\eea

\subsection{Moments of the absorption time, starting at $x$}
Moments of the absorption time are extracted from the Laplace-transformed currents as
\bea\label{Tabs0}
&&\left< T_{\rm exit}^\mu(x)\right>_0 =  -\partial_s \left[ \tilde J_{\rm DD}^\mu(x,1,s) - \tilde J_{\rm DD}^\mu(x,0,s)\right] \Big|_{s=0} \nn\\
&&=    \frac{e^{\mu } (1-x)-e^{\mu  (1-x)}+x}{\left(e^{\mu }-1\right)
   \mu } \\
&& \int_0^1\rmd x  \left< T_{\rm exit}^\mu(x)\right>_0 =  \frac{\mu  \coth \left(\frac{\mu }{2}\right)-2}{2 \mu ^2} \ .
\\
\label{Tabs0squared}
&&\left< T_{\rm exit}^\mu(x)^2\right>_0 =  \partial_s^2 \left[ \tilde J^\mu_{\rm DD}(x,1,s) - \tilde J_{\rm DD}^\mu(x,0,s)\right] \Big|_{s=0} \nn\\
&&\int_0^1\rmd x   \left< T^\mu_{\rm exit}(x)^2\right>_0 = \frac{\mu ^2+3 \mu ^2 \text{csch}^2 (\frac{\mu
   }{2} )-12}{3 \mu ^4}.
\eea

\section{Propagator with one absorbing and one reflecting boundary}
\label{s:prop-one-ab+one-reflecting}
The propagator with an absorbing (Dirichlet) boundary   at $y=0$ and a reflecting (Neumann) one at $y=1$ reads \be\label{PDN}
P_{\rm DN}(x,y,t) = \sum_{n=-\infty}^\infty \frac{(-1)^n }{ \sqrt{4 \pi  t}} \left[ \rme^{-\frac{(2 n+x-y)^2}{4 t}} - 
   \rme^{-\frac{(2 n+x+y)^2}{4 t}} \right].
\ee
The generalization to include drift is as in \Eq{P+mu(x,y,t)}. 
The Laplace transform of \Eq{PDN} is \checked
\bea
\lefteqn{
\tilde P_{\rm DN}(x,y,s)}\nn\\
 &=& \sum_{n=-\infty}^\infty\frac{(-1)^n \left(e^{-\sqrt{s} \left| 2 n+x-y\right| }-e^{-\sqrt{s} \left| 2
   n+x+y\right| }\right)}{2 \sqrt{s}}
 \nn\\
 &=& \frac{\text{sech} (\sqrt{s})}{2
   \sqrt{s}}  \Big[ \sinh \left(\sqrt{s}
   (x+y-1)\right) \nn\\
   && ~~~~~~~~~~~~~~~~~~~-\sinh \left(\sqrt{s} (\left| x-y\right| -1)\right)\Big]\ .
\eea
It can also be written as \checked
\bea
\tilde P_{\rm DN}(x,y,s) &=&   
 \frac{1-\tanh(\sqrt s)}{\sqrt s} \sinh(\sqrt s x)\sinh(\sqrt s y) \nn\\
&& + \frac{e^{-\sqrt{s} \left| x-y\right| }-e^{-\sqrt{s} \left| x+y\right| }}{2 \sqrt{s}} \ .
\eea
Expanding in $s$, we find  
\bea
\tilde P_{\rm DN}(x,y,s) &=&  \min(x,y) - s\left[xy \frac{(x+y)^3-|x-y|^3}{12} \right] \nn\\
&& + ... 
\eea
The outgoing current at the lower boundary is 
\be
\tilde J_{\rm DN}(x,s) =  \frac{\cosh  \big(\sqrt{s} (1-x) \big)}{\text{cosh} (\sqrt{s} )}\ .
\ee
Taylor expanding in $s$ yields
\be
\tilde J_{\rm DN}(x,s) = 1-s \left( x-\frac {x^2}2\right)  + \frac{s^2}{24}  (x^4-4 x^3+8 x) +...
\ee
The first term indicates that all trajectories exist, while the time it takes and its second moment are
\bea
\left< T_{\rm exit}^{\rm DN}(x)\right>  &=& x-\frac{x^2}2\ , \\
\left< T_{\rm exit}^{\rm DN}(x)^2\right>  &=& \frac{x^4-4 x^3+8 x }{12}   \ .
\eea
The propagator with a Dirichlet boundary condition at the upper, and a Neumann boundary condition at the lower end is obtained by replacing $x\to 1-x$ and $y\to 1-y$. 

\section{Propagator with two reflecting boundaries}
\label{s:prop-two-reflecting}
With two reflecting (Neumann) boundary conditions the propagator is 
\bea
P_{\rm NN}(x,y,t) &=& \sum_{n=-\infty}^\infty \frac{1}{ \sqrt{4 \pi  t}} \left[ \rme^{-\frac{(2 n+x-y)^2}{4 t}} + 
   \rme^{-\frac{(2 n+x+y)^2}{4 t}} \right] \nn\\
   &=& \frac{1}{2} \vartheta _3\!\left(\frac{\pi}{2}   (x-y),e^{-\pi ^2
   t}\right)
   \nn\\
   && +\frac{1}{2} \vartheta _3\!\left(\frac{\pi}{2}   (x+y),e^{-\pi ^2
   t}\right)\ .
\eea
Laplace transforming yields
\bea
\tilde P_{\rm NN}(x,y,s) &=&  \frac{\text{csch}  (\sqrt{s})  }{2 \sqrt{s}} \Big[\cosh\!\left(\sqrt{s} (\left|
   x-y\right| -1)\right)\nn\\
   && +\cosh\! \left(\sqrt{s} (x+y-1)\right)\Big]\ .
\eea

\section{Probabilities for  the span}
\label{s:Probabilities-for-the-span}
\subsection{Definition of the span}
The span is a classical problem treated e.g.\ in \cite{Daniels1941,Feller1951,WeissRubin1976,PalleschiTorquati1989}, but the observables we wish to study seem not to have been considered. 
To properly define the problem, we 
note the positive and negative records (the running maximum and minimum) as 
\bea
M_+(t) &:=& \max_{t'\le t}  X_{t'} \ ,\\
M_-(t) &:=& \min_{t'\le t}  X_{t'} \ .
\eea
The span $S(t)$ is their difference, i.e.\ the size of the (compact) domain visited up to time $t$, 
\be
S(t):= M_+(t) - M_-(t)\ .
\ee
(We note capital $S$ for the span, in order to distinguish it from the Laplace variable $s$ conjugate to time $t$.)

\subsection{The probability that the span reaches 1 for the first time}
\label{s:The probability that the span reaches 1 for the first time}
We want to know the probability that $S(t)$ becomes 1 for the first time. We note this time by $T_1$, and its probability distribution by $P_{T_1}(t)$. There are two contributions, depending on whether the process stops while at its minimum or maximum. The probability to stop when the process is at its minimum can be obtained as follows: 
Consider the  outgoing current for the process  starting at $X_0=x$, with  the lower boundary positioned at $m_1$, and the upper boundary  at $m_2$,  i.e.
\bea
\lefteqn{\mathbf J^\mu_{\rm DD}(x,m_1,m_2,t)} \\
&&~ = \frac{1}{(m_2-m_1)^2}\,J_{\rm DD}^{\mu(m_2-m_1)}\!\left( \frac{x-m_{1}}{m_{2}-m_{1}},0,\frac{t}{(m_{2}-m_{1})^{2}}\right). \nn
\eea
(The scale factor can be understood from the observation that the current is a density in the starting point times a spatial  derivative of a probability.)
The probability that the walk reached $m_2$ before being absorbed at $m_1$ is 
$
\partial_{m_2}   {\mathbf J(x,m_1,m_2,t)} 
$. Finally, 
the probability to have span 1 at time $t$ is this expression, integrated over $x$ between the two boundaries.
There is another term, where the process stops while at its maximum. It is obtained from this first contribution when exchanging the  two boundaries, and replacing $\mu$   by $-\mu$. 
Setting w.l.o.g.\ $m_1=0$ and $m_2=m$, the sum of the two terms  is   
\bea\label{PT1-bare}
\lefteqn{P_{T_1}^\mu  (t)=}  \\
&&= - \partial_m \frac1{m^2}\int\limits_0^m \rmd x\, \Big[ J^{\mu m}_{\rm DD}\!\left( \frac{x}{m},0,\frac{t}{m^2}\right) \nn\\
&&\qquad\qquad\qquad\qquad + J_{\rm DD}^{-\mu m}\!\left( \frac{x}{m},0,\frac{t}{m^2}\right)\Big] \bigg|_{m=1}\nn\\
&&= -   \partial_m \frac1{m}\int\limits_0^1 \rmd x\, \Big[ J^{\mu m}_{\rm DD}\!\left( x,0,\frac{t}{m^2}\right)+ J_{\rm DD}^{-\mu m}\!\left( x,0,\frac{t}{m^2}\right) \Big] \bigg|_{m=1} \nn\\
&&=   (1+2 t\partial_t-\mu \partial_\mu)  \int_0^1 \rmd x\, \left[ J_{\rm DD}^\mu\!\left( x,0,t\right) +J_{\rm DD}^{-\mu}\!\left( x,0,t\right) \right]\ .\nn
\eea
For $\mu=0$, this simplifies to 
\bea\label{PT1-bare-mu=0}
P_{T_1}  (t) &=&  2(1+2 t\partial_t)  \int_0^1 \rmd x\, J_{\rm DD}\!\left( x,0,t\right)  \ .
\eea
\begin{figure}
\Fig{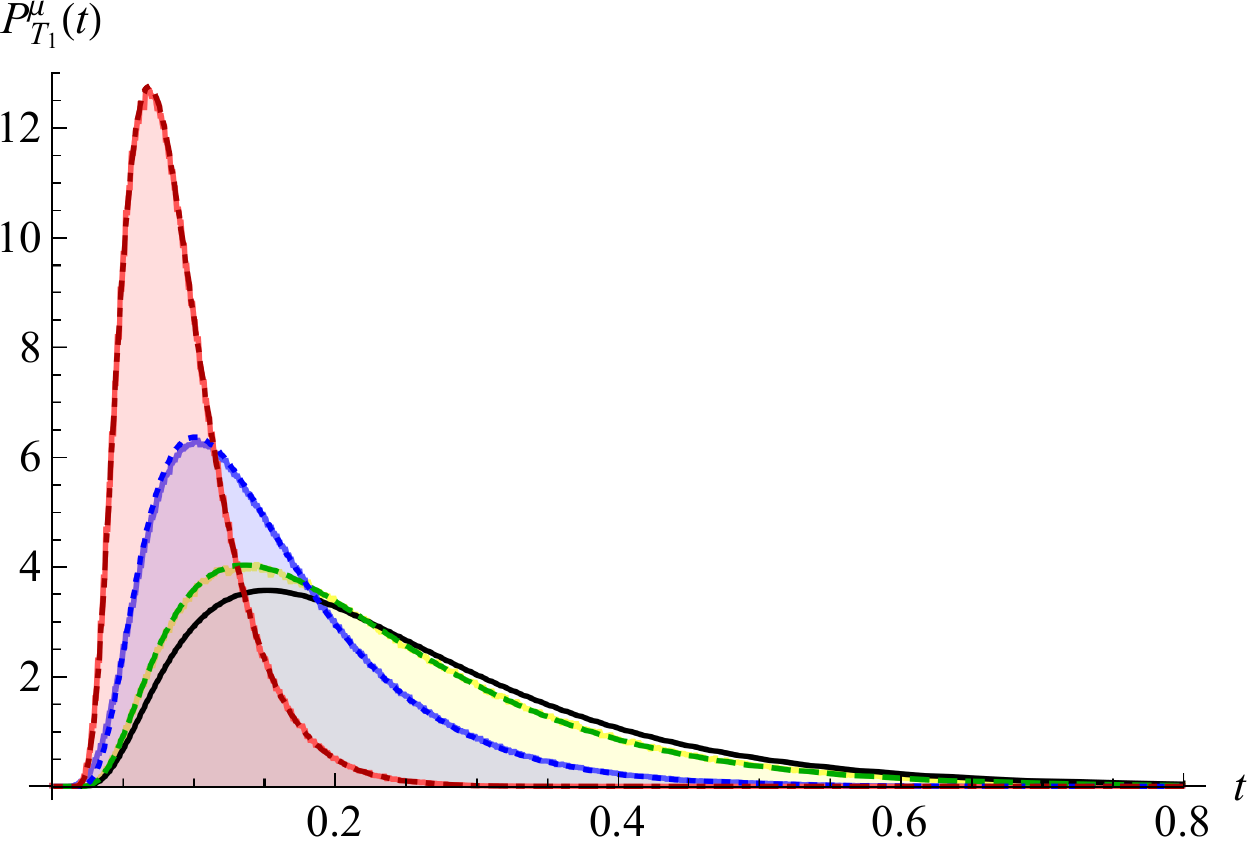}
\caption{The function $P_{T_1}^\mu(t)$ as given in \Eq{PT1mu-ser}, and compared to numerical simulations with $\delta t=10^{-5}$, $10^{6}$ samples. 
Black, solid: $\mu=0$, already shown on Fig.~\ref{f:Pspan}. Green dashed: $\mu=2$. Blue dotted: $\mu=5$. Red, dash-dotted: $\mu=10$.}
\end{figure}Using Eqs.~(\ref{22}) 
and  (\ref{P+mu(x,y,t)}) 
allows us to rewrite the integral (for $\mu=0$) as 
\bea
\lefteqn{ \int\limits_0^1 \rmd x\, J_{\rm DD}\!\left( x,0,t\right)} \nn\\
&=&\int\limits_0^1 \rmd x\, \partial_y \left[ \mathbb P(x-y,t) -\mathbb P(x+y,t) \right] \bigg|_{y=0} \nn\\
 &=&-2 \int\limits_0^1 \rmd x\, \partial_x   \mathbb P(x,t)   = 2\left[  \mathbb P(1,t) -  \mathbb P(0,t) \right] \ .  \qquad\qquad\qquad
 \eea
\pagebreak[1]
Thus 
\be\label{47}
P_{T_1}(t) =  4 (1+2 t \partial_t)  \left[  \mathbb P(1,t) -  \mathbb P(0,t) \right]\ .
\ee\begin{figure*}[t]
{\fig{8cm}{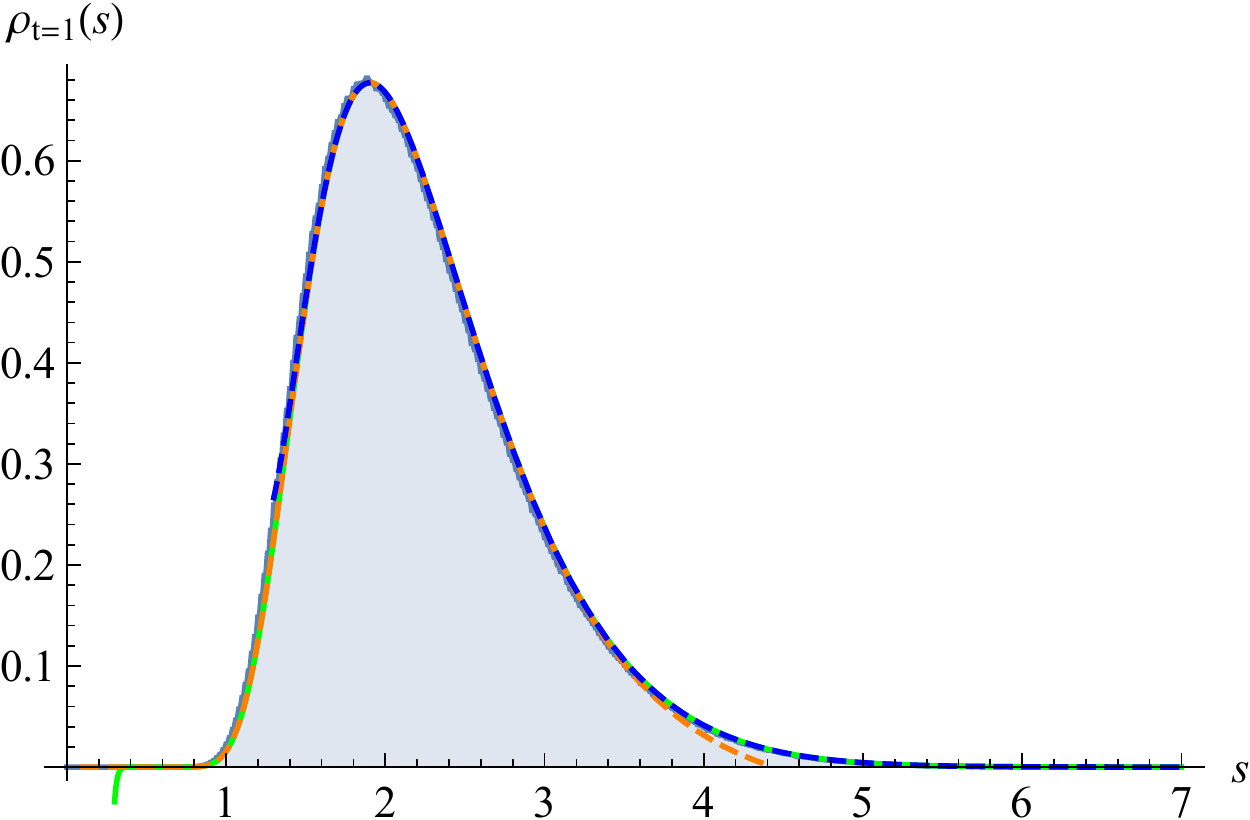}}~~~~~~~~~{\fig{8cm}{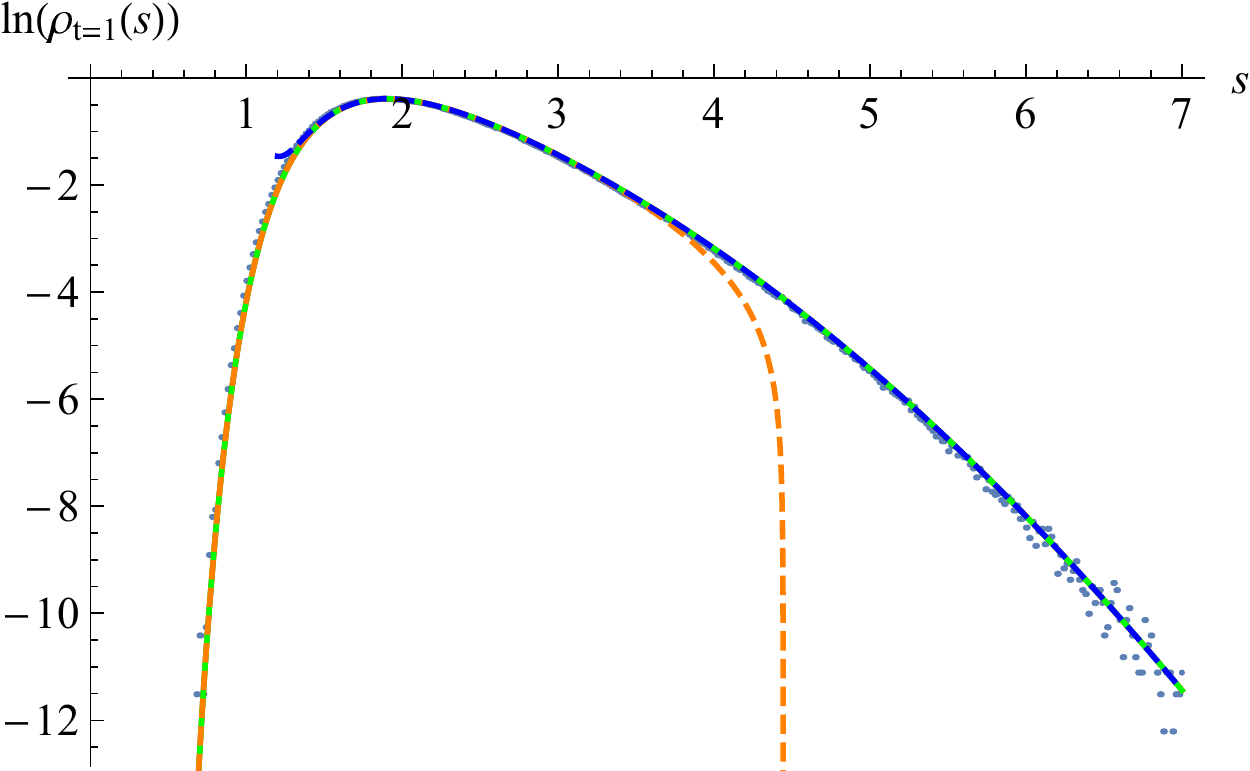}}
\caption{Left: The density of the span at time $t=1$. Grey: RW simulation with $\delta t=10^{-4}$, and $10^6$ samples. Green: the analytic result (\ref{rho-rhospan-ana}). 
Orange dashed the small-$s$ asymptotics (\ref{rho-small}); blue dashed the large-$s$ asymptotics (\ref{rho-large}). Right: {\em ibid.} on a log-scale.}
\label{f:rhospan}
\end{figure*}
\begin{figure}[t]
\vspace*{-0.2cm}
\Fig{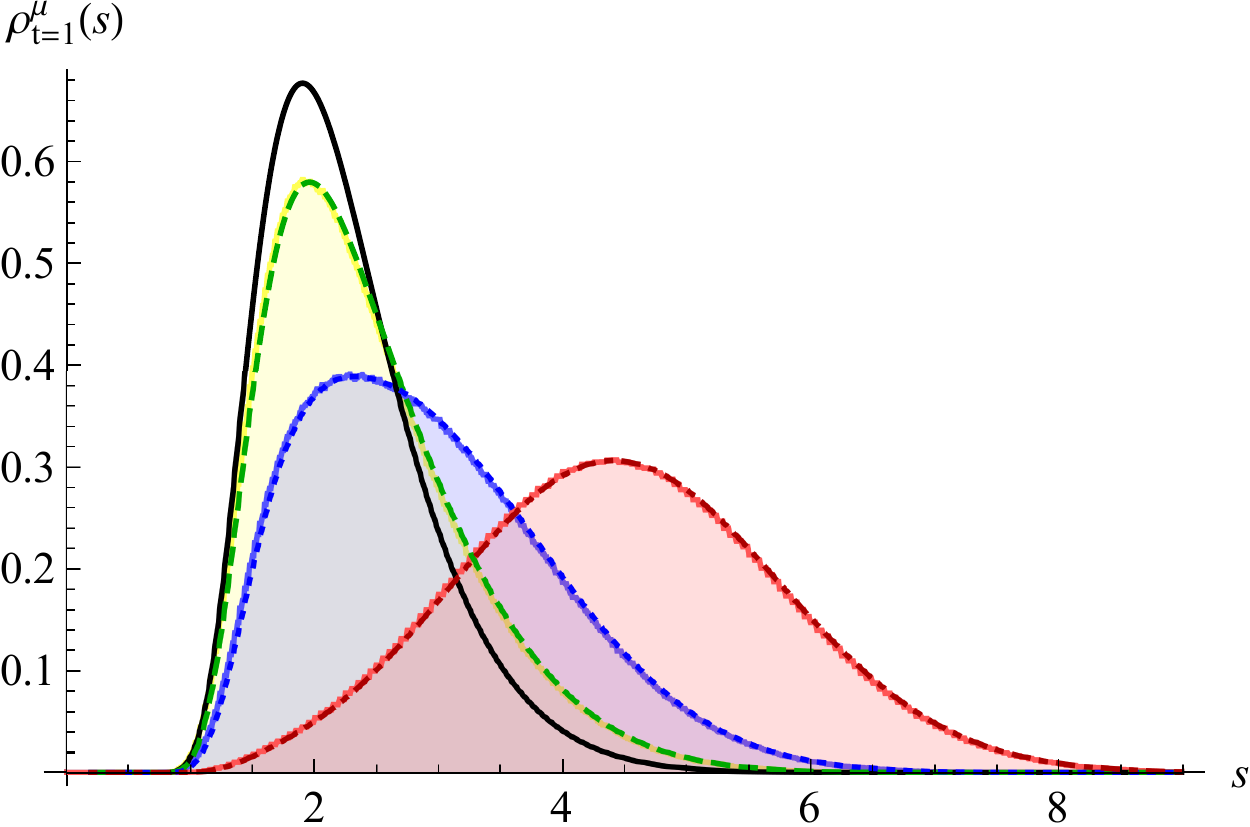}
\caption{The density $\rho_{t=1}^\mu(s)$, for $\mu=0$ (black solid line, with data only  shown on Fig.~\ref{f:rhospan}), $\mu=1$ (green dashed line, with data in yellow), $\mu=2$ (blue dotted line, data in light blue), and $\mu=4$ (red dot-dashed line, data in light red). Numerical validation  with $\delta t=10^{-5}$, and $10^{6}$ samples.}
\label{f:rhosmu}
\end{figure}
Inserting the definition \eq{mathbbP} of $\mathbb P$, we get
\bea\label{PT1}
P_{T_1}(t) &=& 4  \big(  1 +2 t  \partial_{t} \big) \sum_{n=-\infty}^\infty \frac{e^{-\frac{(2 n+1)^2}{4 t}} - e^{-\frac{n^2}{t}}}{\sqrt{4\pi t}}
 \nn\\
 &=& \frac1 {\sqrt{\pi } t^{3/2} }\sum_{n=-\infty}^\infty   (2 n+1)^2 e^{-\frac{(2 n+1)^2}{4 t}}-4 n^2 e^{-\frac{n^2}{t}} \nn \\
   &=& 4 \sqrt{\frac t \pi } \partial_t \left[  \vartheta _2 \!\left(0,e^{-1/
   t}\right)- \vartheta _3 \!\left(0,e^{-1/
   t}\right)\right] \ .
\eea
With the help of the Poisson-formula transformed Eq.~(\ref{PP-Poisson}) this can   be written as
\be\label{PT1-Poisson}
P_{T_1}(t) = 8 \sum_{n=0}^\infty \rme^{-\pi^2 (2n+1)^2 t}\left[ 2 \pi^2 (2n+1)^2 t-1 \right] \ .
\ee
This result is compared to a numerical simulation on Fig.~\ref{f:Pspan}. 
Our expansions allow us to give simple formulas for the small and large-$t$ asymptotics, 
\bea\label{PT1-small}
P_{T_1}(t) &\simeq& \frac{2\rme^{-\frac1{4t}}}{\sqrt {\pi }t^{3/2}} + \ca O(\rme^{-\frac1{t}} )\ , \\
\label{PT1-asymp}
P_{T_1}(t) &\simeq&   \rme^{-\pi^2 t} \left[16 \pi^2 t  -8    +\ca O(e^{-8 \pi^2 t}) \right]\ .
\eea 
These expansions work in a rather large, and overlapping domain, as can be seen on Fig.~\ref{f:Pspan}.
Its Laplace transform is \bea\label{tildePT1}
\tilde P_{T_1}(s) &=& 2(1+2 s \partial_s) \sum_{n=-\infty}^{\infty} \int_0^\infty \rmd t\, \frac{   e^{-\frac{n^2}{t}} - \rme^{\frac{(2 n+1)^2}{4 t}}}{\sqrt{\pi t}}\, \rme^{-st }
\nn\\
&=& \frac1{\cosh(\sqrt s/2)^2}\ .
\eea
Extracting the moments from the Laplace transform   yields
\be
\left< T_1\right> =\frac14\ , \quad \left< T_1^2\right> =\frac1{12} \ , \quad \left< T_1^3\right> = \frac{17}{480}\ ,\quad...
\ee

Let us now return to the case with drift in \Eq{PT1-bare}.  
Since formulas become rather cumbersome, we only give one well-converging series expansion, based on the   representation \eq{PP-Poisson},
\bea\label{PT1mu-ser}
&&P_{T_1}^\mu  (t)= \sum_{n=1}^\infty a_n^\mu + a_n^{-\mu} \\
&&a_n^\mu = \frac{4 \pi ^2 n^2 e^{-\pi ^2 n^2 t-\frac{1}{4} \mu  (\mu 
   t+2)}}{\left(\mu ^2+4 \pi ^2 n^2\right)^2} \times  \\
   &&\times \bigg[ 2 e^{\mu /2} \Big(8 \pi ^4 n^4 t+2 \pi ^2 n^2
   (\mu ^2 t-2) -3 \mu ^2 \Big) \nn\\
   && +(-1)^n \Big(\mu ^2 (\mu +6)-16 \pi ^4 n^4 t+4 \pi ^2
   n^2 (\mu -\mu ^2 t+2)\Big) \bigg]\nn . 
\eea

\subsection{Density of the span}
Let us connect to the classical work on the span \cite{Daniels1941,Feller1951,WeissRubin1976,PalleschiTorquati1989}. We will show how to reproduce formulas (3.7)-(3.8) in \cite{Feller1951}.
The latter give the density $\rho_t(S)$ for the span $S$ at time $t$. 
In our formalism, it can be obtained as 
\be\label{59}
\rho_t^\mu(m_2-m_1) =-\partial_{m_1}\partial_{m_2} \int\limits_{m_1}^{m_2}\!\rmd x  \int\limits_{m_1}^{m_2}\!\rmd y\, \mathbf P^\mu_{\rm DD}(x,y,m_1,m_2,t), 
\ee
where $\mathbf P^\mu_{\rm DD}(x,y,m_1,m_2,t)$ is the probability to go from $x$ to $y$ in time $t$, without being absorbed by the lower boundary positioned at $m_1$, or the upper boundary positioned at $m_2$.
In terms of the propagator $P_{\rm DD}^\mu(x,y,t)$, this can be written as 
\be\label{60}
\rho_t^\mu(S) =\partial_{S}^2 \left[  S\int\limits_{0}^{1}\rmd x\,  \int\limits_{0}^{1}\rmd y\, P^{\mu S}_{\rm DD}(x,y,t/S^2)\right] \ .
\ee
We start with $\mu=0$: 
Using \Eq{P+mu(x,y,t)}, and the series expansions \eq{mathbbP} and \eq{PP-Poisson} yields after integration and simplifications two different representations, 
\bea
\rho_t(S) &=& \frac{4}{\sqrt{\pi t}} \sum_{n=1}^\infty  (-1)^{n+1} n^2 e^{-\frac{n^2 S^2}{4 t} } \nn \\
&=& \frac{16 t}{S^3} \sum_{n=0}^\infty  e^{-\frac{\pi ^2   (2 n+1)^2 t}{S^2}} \Big[\frac{2 \pi ^2 (2 n+1)^2 t}{S^2} -1\Big]\ .\qquad 
\label{rho-rhospan-ana}
\eea
This is equivalent to Eqs.~(3.7)-(3.8) in \cite{Feller1951}, if one there replaces $t\to 2 t$. (Our variance \eq{variance} is $2t$ instead of $t$ as in \cite{Feller1951}.) 
The small and large-$S$ asymptotics are \bea\label{rho-small}
\rho_t(S) &\simeq& \frac{4}{\sqrt{\pi t}}  \left[  e^{-\frac{S^2}{4t} } - 4 e^{-\frac{S^2}{t} }  + \ca O\Big( e^{-\frac{9 S^2}{4t} } \Big)  \right] \ ,\\
\rho_t(S) &\simeq& \frac{16 t}{S^5}   e^{-\frac{\pi ^2  t}{S^2}} \Big[2 \pi ^2 t-S^2\Big] +\ca O\Big( s^2\rme^{-\frac{9 \pi^2 t}{S^2}}\Big) \ .\qquad 
\label{rho-large}
\eea
Note that in \Eq{rho-small} we have also retained the subleading term for small $S$, which considerably improves the numerical accuracy.  A test is presented on Fig.~\ref{f:rhospan}.

Let us now turn to the general case with $\mu \neq 0$. 
There we have using the generating function \eq{PP-Poisson}
\begin{widetext}\vspace*{-1mm}
\bea
\rho_t^\mu(S)
   &=&  \partial_S^2  \sum_{n=1}^\infty\frac{32 \pi ^2 (-1)^{n+1} n^2 S \left[(-1)^{n} e^{\frac{\mu 
   S}{2}}-1\right]^2 e^{-\frac{\pi ^2 n^2 t}{S^2}- \frac{1}{4}
   \mu  (2 S+\mu  t)}}{\left(4 \pi ^2 n^2+\mu ^2 S^2\right)^2}\ .
\eea
This formula is checked on Fig.~\ref{f:rhosmu}. 
The small-$S$ asymptotics can be obtained by retaining only the leading term in $n$. 
Let us finally note that for large $\mu$, this density tends to 
\be
\rho_t^\mu(S) \to \frac{1}{\sqrt{{4\pi t}}} \,\left[  \rme^{-\frac{(S-\mu  t)^2}{4 t}}+  \rme^{-\frac{(S+\mu  t)^2}{4 t}}\right] \ , \qquad |\mu|  \gg 0 \ .
\ee
\begin{figure*}
\fig{8.5cm}{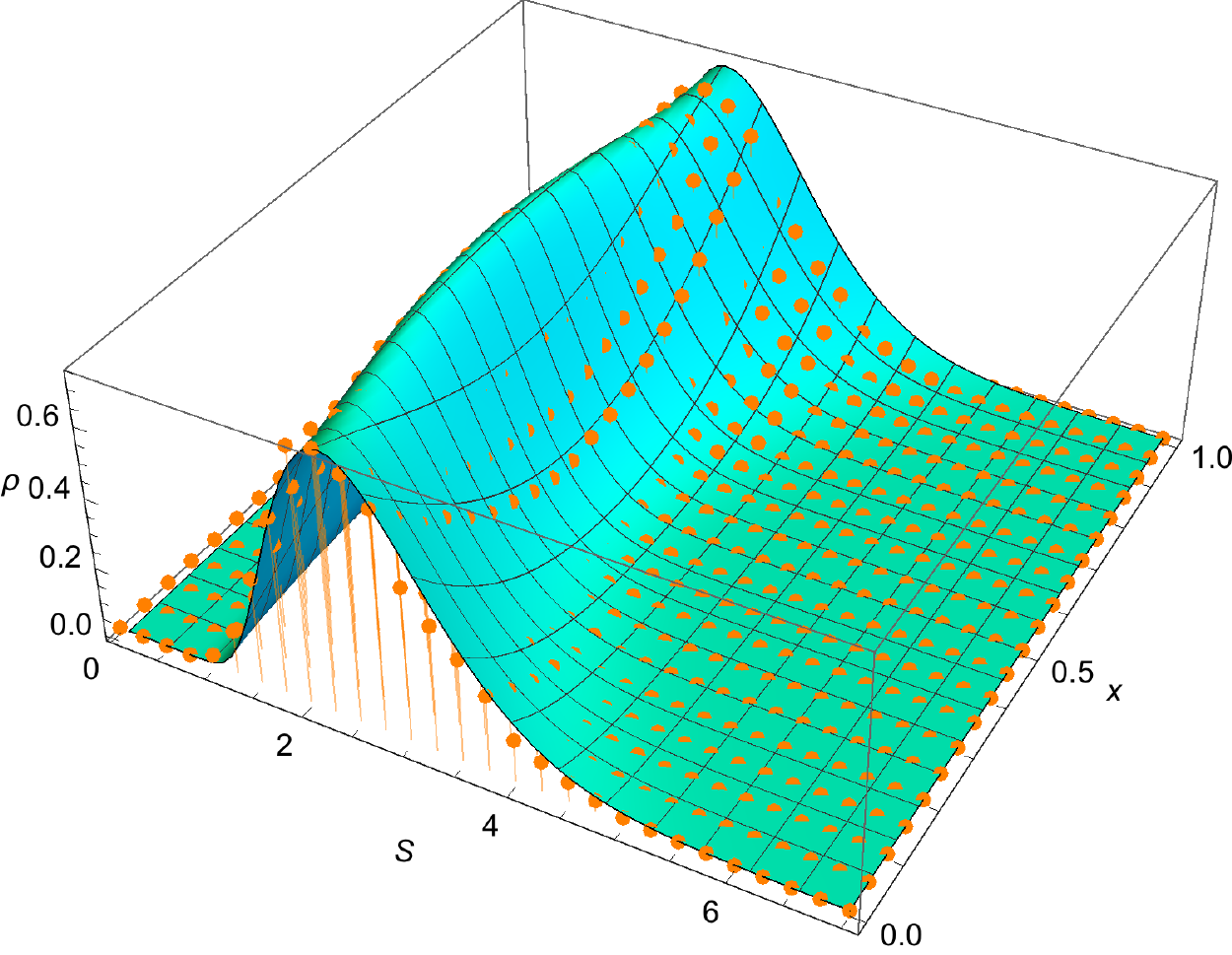}\hfill\fig{8.5cm}{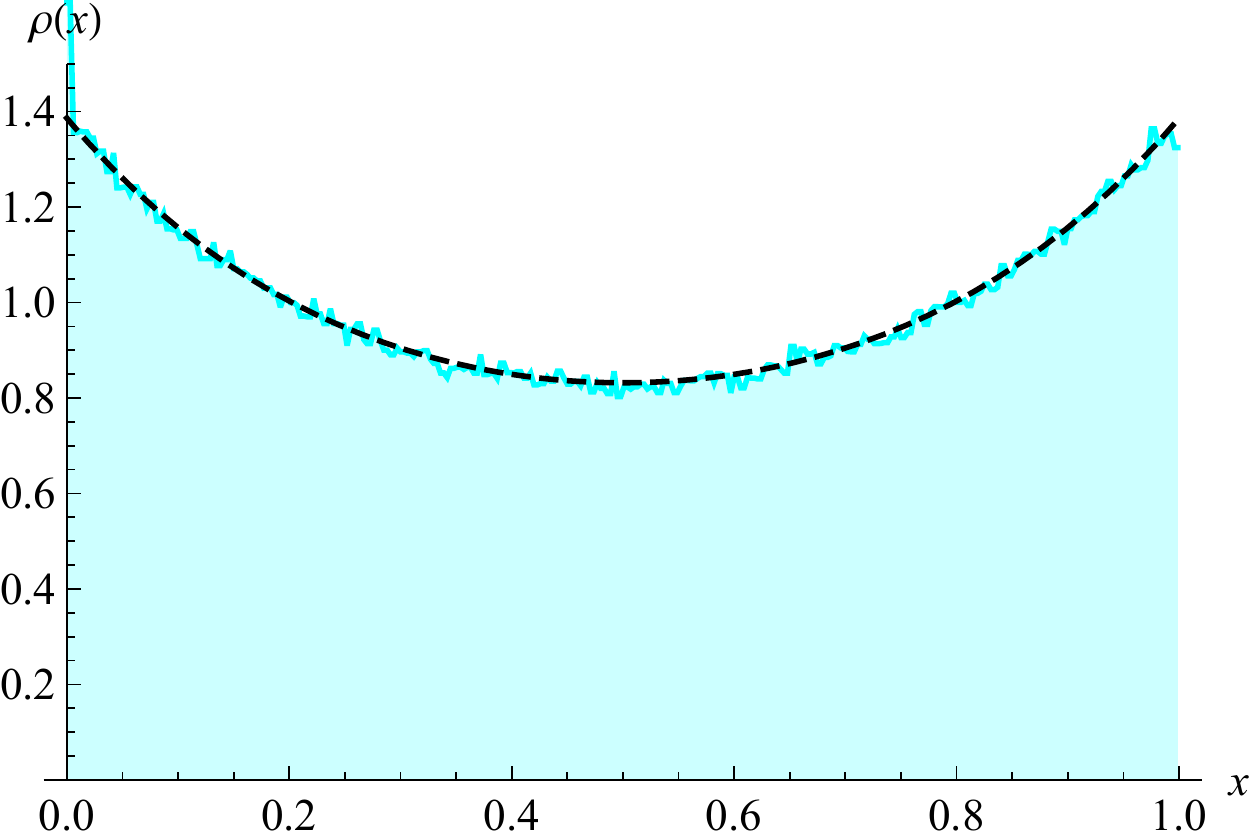}
\caption{Left: The density $\rho_{t=1}(S,x)$ as given in \Eq{rho-max-min}. Right: The density $\rho(x)$ as given in \Eq{71}. The numerical validation (left: orange dots; right: cyan shaded region) was performed with $\delta t=10^{-5}$, and $10^{6}$ samples.}
\end{figure*}For $\mu>0$, 
the first term is the probability density for the max of the endpoint, supposing that the minimum is at $0$. For $\mu<0$, the second term arises, with max and min interchanged.   
\end{widetext}

\section{Joint density of maximum and minimum}
\label{s:joint-max+min}
We can also derive the joint density of the maximum $M_{+}\equiv m_2>0$ and minimum $M_{-}\equiv m_{1}<0$, starting at $x=0$. 
In analogy of \Eq{59}, this can be written as 
\be\label{67bis}
\rho_t^\mu(m_2,m_1) =-\partial_{m_1}\partial_{m_2}  \int\limits_{m_1}^{m_2}\rmd y\, \mathbf P^\mu_{\rm DD}(0,y,m_1,m_2,t)\ . 
\ee
\pagebreak[9]
The equivalent of \Eq{60} then becomes
\bea
&&\rho_t^\mu(m_2,m_1)  \\
&& =-\partial_{m_{1}}\partial_{m_{2}}  \int\limits_{0}^{1}\rmd y\, P^{\mu (m_{2}-m_{1})}_{\rm DD}\left(\textstyle\frac{-m_{1}}{m_{2}-m_{1}},y,\frac{t}{(m_{2}-m_{1})^{2}} \right)   . \nn
\eea
Inserting \Eq{P+mu(x,y,t)} and one of the two representations \eq{mathbbP} or \eq{PP-Poisson} yields two converging series expansions. Since in general these expressions are little enlightening, we continue with $\mu=0$. 
To simplify our analysis, we rewrite the density \eq{67bis} in terms of $S:=m_{2}-m_{1}$ and $x:=m_{1}/(m_{1}-m_{2})$:
\be
\rho_{t}(x,S) := S \rho_{t}^{0}\big(-x S,(1-x)S \big)\ .
\ee
Its marginal density coincides with   \Eq{rho-rhospan-ana}\ , 
\be
\int_{0}^{1}
\rho_{t}(x,S) \,\rmd x = \rho_{t}(S)\ .
\ee
\begin{widetext}\noindent
The two series expansions in question are \bea\label{rho-max-min}
\rho_{t}(S,x)  &=& \frac{4 }{S}\sum_{n=0}^{\infty} e^{-\frac{\pi ^2 (2n+1)^2 t}{S^2}}
\bigg[  \cos \!\Big(  (2n+1) x \pi \Big) (2 x-1) \left(1-\frac{2 \pi ^2  (2n+1)^2 t}{S^{2}}\right)\nn\\
&&\qquad\qquad\qquad\qquad~  +\pi  (2n+1) \sin\!
   \Big(   (2n+1) x \pi\Big) \left(\frac{4 \pi ^2  (2n+1)^2 t^2}{S^{4}} +  x (1-x) -6  \frac t {S^{2}}\right) \bigg]\nn\\
&=&\frac1{\sqrt{\pi t}} \,\partial_{x} \sum_{n=-\infty}^{\infty} (-1)^n n (n+1) e^{-\frac{S^2 (n+x)^2}{4 t}} \ .
\eea
Interestingly, the latter equation allows us to obtain the marginal distribution of $x$ in closed form. Since this function is independent of $t$, we drop the time index:
\bea\label{71}
\rho(x) &:=& \int_{0}^{\infty} \rmd S\, \rho(x,S) = \partial_{x} \sum_{n=-\infty}^{\infty} (-1)^n \frac{n (n+1)}{|n+x|}\nn\\
&=& 1+\partial_{x}  \left\{ \frac{x(1-x)}{2} \left[\psi
   \left(\frac{2-x}{2}\right)-\psi
   \left(\frac{3-x}{2}\right)-\psi
   \left(\frac{x+1}{2}\right)+\psi
   \left(\frac{x+2}{2}\right)\right]\right\}\ .
\eea
\end{widetext}
This  is the density for the relative  location of the starting point w.r.t.\ the domain given by the maximum and  minimum. It is also the distribution of the final position w.r.t.\ the same domain. This density is larger at the boundaries, as is easily understood: After a new record, the particle diffuses away from the record, but the probability density remains higher close to the last record.

\section{The span with a reflecting wall}
\label{s:Diffusion with a reflecting wall}

Now consider diffusion with   a reflecting wall at $x=0$. We want to know the probability density for the span to reach $1$ for the first time. For simplicity, we restrict to the drift-free case $\mu=0$. We also assume $x<1$, since  for $x>1$ the reflecting boundary can never be reached, and we recover the result of section \ref{s:The probability that the span reaches 1 for the first time}.
Suppose the process starts at $x$, with $0\le x<1$. There are two possibilities: Either the process first reaches  0, or   1. The probabilities for these two events are $x$ and $1-x$, respectively. If it first reaches 1, then it almost surely also reaches $1+\delta$ with $\delta$ small before its span becomes 1; as a consequence its minimum is bounded by $\delta$. Thus it   never reaches the lower boundary at $x=0$.

Consider the two contributions in turn: The first contribution, when the process never reaches $x=0$, is similar to the one obtained in Eq.~\eq{PT1-bare}. It can itself be decomposed into two sub-contributions, depending on whether, when the span reaches 1, $X_t$ equals its maximum (case 1a) or minimum (case 1b). We start with case 1a. Denoting $\mathbf J_{\rm DD}(y,m_2,t|m_1,m_2)$ the outgoing current  at the upper boundary $m_2$, for a particle starting at $y$, with lower boundary $m_1$, 
we have 
\bea\label{P1} \nn
p_{1a} (x,t)&=& -   \int\limits_0^x \rmd y\,   \partial_{m_1}  \mathbf J _{\rm DD}(y,m_2,t|m_1,m_2)  \bigg|_{m_2=m_1+1}\nn\\
&=& -\int\limits_0^x \rmd y\,   \partial_{m_1}  \bigg[\frac1{(m_2{-}m_1)^2} \nn\\
&& \qquad \times J_{\rm DD}\!\left(\frac{y{-}m_1}{m_2{-}m_1},1,\frac t{(m_2{-}m_1)^2}\right)  \bigg]_{m_1=0}^{m_2=1}. \nn\\
\eea
Let us first evaluate its normalization, using that  the time-integrated current is the exit probability,
\bea\label{65}
\int_0^\infty \rmd t\, p_{1a}  (x,t)&=& - \int_0^x\rmd y\, \partial_{m_1} \frac{y-m_1}{m_2-m_1} \bigg|_{m_1=0}^{m_2=1} \nn\\
&=&  \int_0^x\rmd y\,(1-y) = x - \frac{x^2}2\ .
\eea
Note that this is smaller than the probability $x$ to exit at the upper boundary. This can be understood from the fact that the trajectory has   to go beyond  1, 
\begin{figure*}
\fig{6cm}{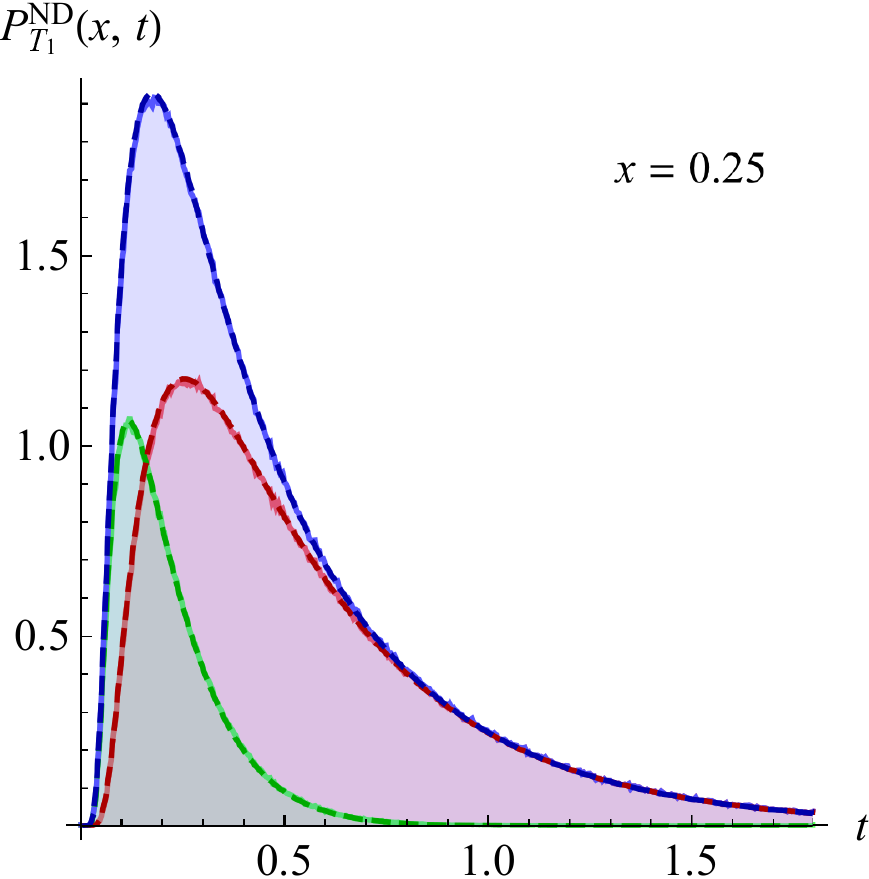}\fig{6cm}{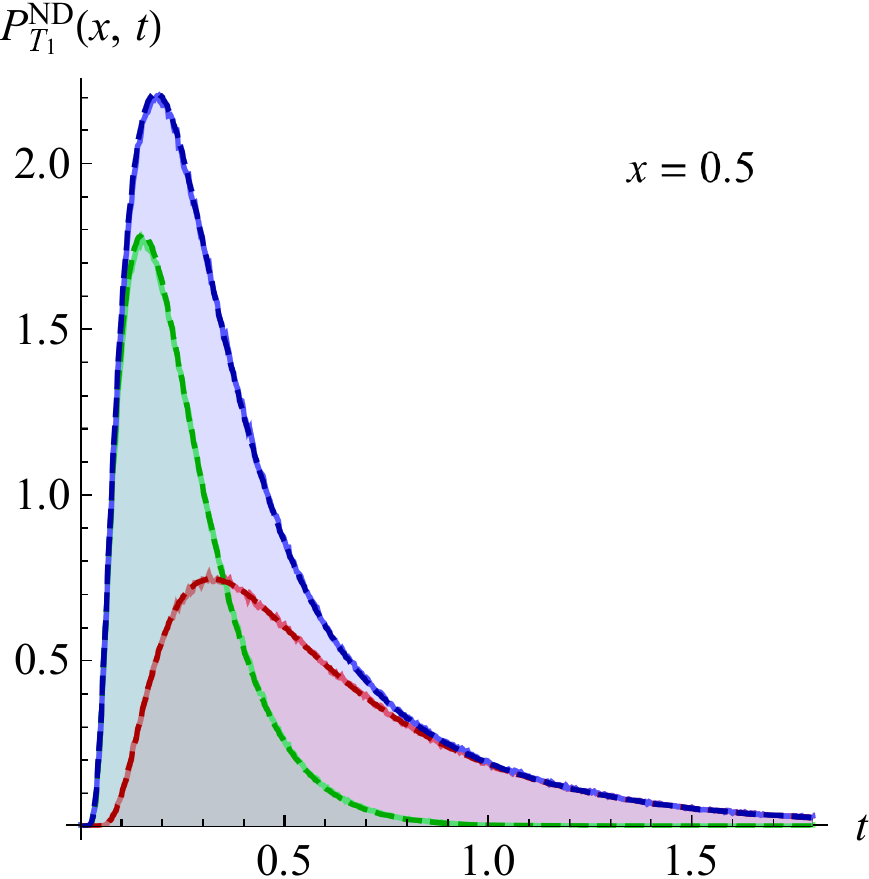}\fig{6cm}{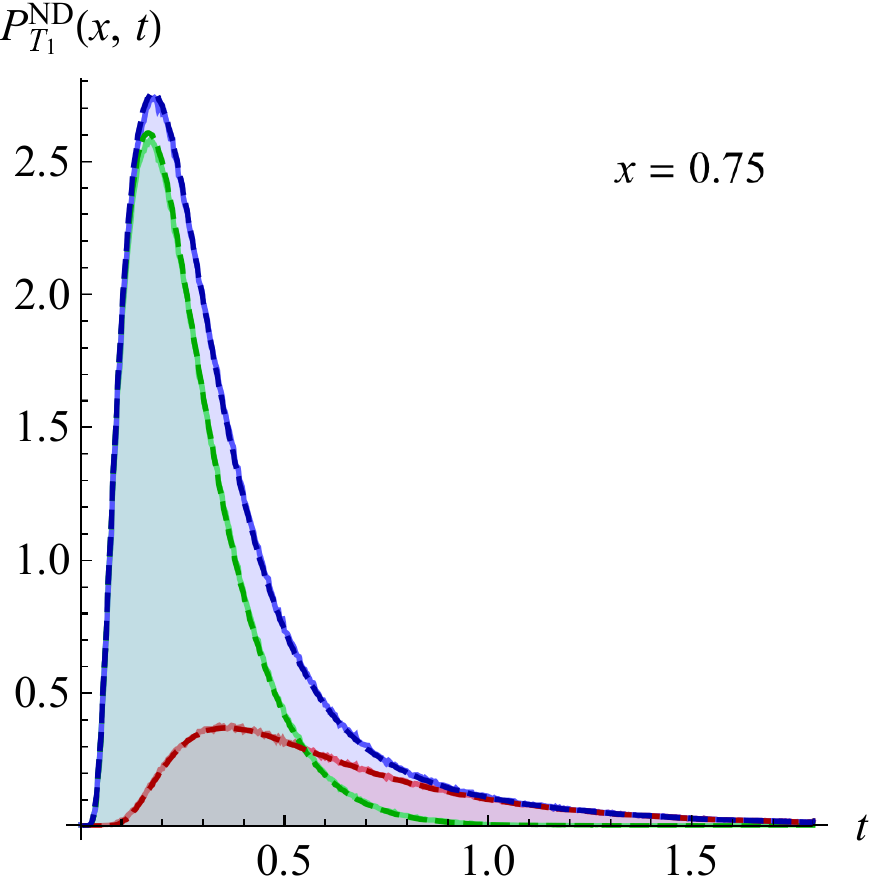}
\caption{The probability $P_{T_1}(t)$ with a reflecting boundary at 0, and an absorbing one at $1$, for $x=0.25$, $x=0.5$, and $x=0.75$.
The green contribution is $p_1(x,t)$, while the red one is  $p_2(x,t)$. Their sum is given in blue. The dark dashed lines are the analytic curves, while the numerical data are given as shaded regions with their envelope in the same color.  $10^7$ samples where simulated, with a time-step of $\delta t=10^{-5}$.
}
\label{f:PT1-reflect}
\end{figure*}
or more precisely to $1+\rm min$, where ${\rm min}>0$ is the minium of the trajectory. 
Continuing with \Eq{P1}, we obtain 
\bea
p_{1a} (x,t) = 2\int_0^x \rmd y \, \Big[ &-&2\partial_y  \mathbb P(1-y,t) \nn\\
& -& 2\partial_y\partial_t  \mathbb P(1-y,t) \nn\\&+&(1-y)\partial_y^2  \mathbb P(1-y,t)\Big]\ .
\label{67}
\eea
Integrating this yields
\begin{align}
p_{1a} (x,t) &=  2 \Big[ \mathbb P(1,t)-\mathbb P(1-x,t)\Big] \nn\\
&+ 4 t \partial_t \Big[\mathbb P(1,t)-\mathbb P(1-x,t) \Big] \nn\\
& +2 \partial_y\mathbb P(y,t)|_{y=1} + 2 (1-x) \partial_x \mathbb P(1-x,t)  \ .
\end{align}
(The first term on the last line vanishes). 
To simplify this expression, introduce the function $\mathbb R(x,t)$ defined as 
\be
\mathbb R(x,t) := -2 \Big[  1+ 2t \partial_t -(1-x)\partial_x \Big]\mathbb P(1-x,t)\ .
\ee
This yields
\be\label{70}
p_{1a}(x,t) = \mathbb R(x,t)-\mathbb R(0,t)\ .
\ee
This is written s.t.\ $\mathbb R$ can be thought of as the principle function of the integrand in \Eq{67}.
The second contribution   where the process never reaches 0 is obtained when the process has its maximum at $1+\delta$ with $\delta >0$, before going down to $\delta<x$, where the process stops (case 1b). 
By symmetry, this is the same expression as \Eq{P1}, where all positions $x$ are sent to $1-x$, i.e.\ 
\bea\label{P1b} \nn
p_{1b}  (x,t)&=& -   \int\limits_{1-x}^1 \rmd y\,   \partial_{m_1}  \mathbf J  _{\rm DD}(y,m_2,t|m_1,m_2)  \bigg|_{m_2=m_1+1}\\
&=& -\int\limits_{1-x}^1 \rmd y\,   \partial_{m_1}  \bigg[\frac1{(m_2{-}m_1)^2} \nn\\
&& \qquad \times J_{\rm DD}\!\left(\frac{y{-}m_1}{m_2{-}m_1},1,\frac t{(m_2{-}m_1)^2}\right)  \bigg]_{m_1=0}^{m_2=1} .\nn\\
\eea
The probability for this  process is as in \Eq{65} given by the time-integrated current  \bea
\int_0^\infty \rmd t\, p_{1b} (x,t)&=& - \int_{1-x}^1 \rmd y\, \partial_{m_1} \frac{y-m_1}{m_2-m_1} \bigg|_{m_1=0}^{m_2=1} \nn\\
&=&  \int_{1-x}^1\rmd y\,(1-y) = \frac{x^2}2\ .
\eea
Thus, as expected
\be
\int_0^\infty \rmd t\, p_{1}  (x,t)= \int_0^\infty \rmd t\,
p_{1a}  (x,t)+p_{1b}    (x,t)= x\ .
\ee
Let us  continue with the evaluation of $p_{1b}(x,t)$, 
\bea
p_{1b}  (x,t) = 2\int_{1-x}^1 \rmd y \, \Big[ &-&2\partial_y  \mathbb P(1-y,t) \nn\\
& -& 2\partial_y\partial_t  P(1-y,t) \nn\\&+&(1-y)\partial_y^2  P(1-y,t)\Big]\ .
\eea
Integration yields
\begin{align}
p_{1b} (x,t) =& \, 2 \Big[ \mathbb P(x,t)-\mathbb P(0,t)\Big] \nn\\
&+ 4 t \partial_t \Big[\mathbb P(x,t)-\mathbb P(0,t) \Big] \nn\\
& + 2 x \partial_x \mathbb P(x,t)  \ .
\end{align}
In analogy of \Eq{70} this can be written as 
\be
p_{1b} (x,t) = \mathbb R(1,t)-\mathbb R(1-x,t)\ .
\ee
The sum of the two  contributions  $p_{1a}$ and $p_{1b}$ is
\bea
&&p_{1}  (x,t)  = p_{1a}    (x,t) +p_{1b}     (x,t) \nn\\
&&~~= 2(1+2t \partial_t ) \left[ \mathbb P(1,t)-\mathbb P(0,t)+\mathbb P(x,t)-\mathbb P(1-x,t)\right] \nn\\
&&~~~~~~~+2 x \partial_x \mathbb P(x,t)+2 (1-x)\partial_x \mathbb P(1-x,t) \nn\\
&&= \mathbb R(x,t) -\mathbb R(0,t)-\mathbb R(1-x,t) +\mathbb R(1,t)\ .
\eea
Note that for $x=\frac12$, one gets $p_{1}  (x,t) =\frac12 P  _{T_1}(t)$.

The second contribution is achieved when the process first reaches the lower boundary. It can be obtained 
by folding the probability to first reach the lower boundary, i.e. the outgoing current at $x=0$, with an absorbing boundary both at $x=0$ and $x=1$,  with the outgoing current at $x=1$ with a reflecting boundary   at $x=0$ and an absorbing one at 1, i.e.
\be
 p_{2}  (x,t) =-\int_0^t \rmd \tau \, J_{\rm DD}  (x,0,\tau)  J  _{\rm ND}(0,1,t-\tau)\ . \ee
Passing to Laplace variables, this reads
\be\label{75}
\tilde p_{2}  (x,s) =- \tilde J_{\rm DD}  (x,0,s) \tilde J  _{\rm ND}(0,1,s) \ .
\ee
We had calculated the currents before, 
\bea
- \tilde J_{\rm DD}(x,0,s) &=&\frac{ \sinh \big(\sqrt {s}  (1-x)\big) }{\sinh(\sqrt {s} )} \ , \\
\tilde J_{\rm ND}(0,1,s) &=& \frac1{\cosh(\sqrt{s})}\ .
\eea
The inverse Laplace transform of \Eq{75} can be written as 
\bea\label{78}
 p_{2}   (x,t) &=& \sum_{n=-\infty}^\infty \frac{\rme^{-\frac{(1-4n+x)^2}{4t}} (1-4n+x)}{\sqrt{\pi }t^{3/2}}\nn\\
 &=& -\partial_x\, \vartheta _3\bigg(\!\!-\frac{\pi}{4}  
   (x+1),e^{-\frac{\pi ^2
   t}{4}}\bigg) \nn\\
   &=& -2 \,\partial_x\, {\mathbb P}\bigg(\frac{1+x}2,\frac t 4\bigg) \ .
\eea
This is checked by evaluating the Laplace transform of each term in the above sum, and then performing the sum over $n$. 

The probability to first reach the lower boundary is  
\be
\int_0^\infty \rmd t\, p_2(x,t)=1-x\ .
\ee

The probability to reach span 1, starting at $x$, and with a reflecting boundary at $x=0$ is finally obtained as 
\be
P_{T_1}^{\rm ND}(x,t) =  p_{1} (x,t)+ p_{2}  (x,t)\ .
\ee
The mean time to reach span 1 is 
\be
\left< T_1^{\rm ND}(x)\right> =\frac{1}{2}-\frac{x^2}{4} \ .
\ee
Thus when starting close to the reflecting wall, it takes  on average twice as long to reach span 1, as  when   starting from far away. 

A numerical check for $x=0.25$, $x=0.5$, and $x=0.75$ is presented on Fig.~\ref{f:PT1-reflect}.

\section{The span with two reflecting boundaries}
\label{s:The span with two reflecting boundaries}
Finally, consider two reflecting (Neumann) boundaries at $x=0$ and $x=a\ge 1$, and suppose that $0<x<1$, and $0<a-x<1$, so that both boundaries can be reached before the span attains one and the process terminates. These conditions can be summarized in 
 \be\label{x-cond}
 a-1<x<1 \ .
 \ee
In   generalization of \Eq{78}, one can write\begin{figure}[t]
\Fig{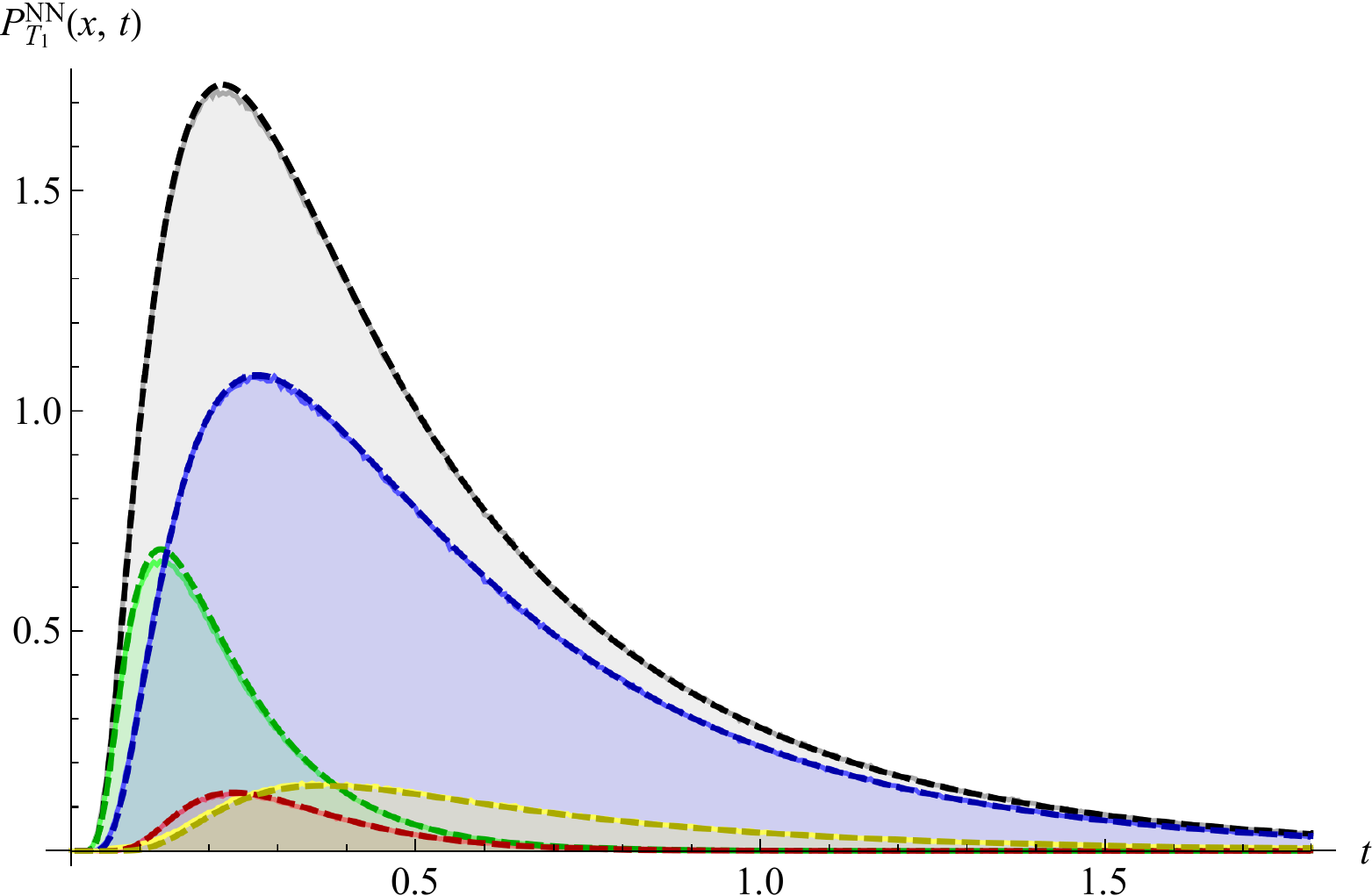}
\caption{The probability $P^{\rm NN}_{T_1}(x,t)$, for $x=0.3$, $a=1.2$. The contributions  are  $p_2(x,t)$ (blue), $p_2(a-x,t)$ (yellow), $p_3(x,t|a)$ (green), and $p_3(a-x,t|a)$ (red).}
\label{f:PT1NNweighted}
\end{figure}\bea\label{88}
 P_{T_1}^{\rm NN}(x,t) &=& p_2(x,t)+p_2(a-x,t) \nn\\
 &&  + p_{3} (x ,t |a)+  p_{3} (a-x ,t|a)  \ . \qquad 
\eea 
The function $ p_{3} (x ,t |a)$ is a modification of $p_{1a} (x ,t )$, defined by 
\bea
p_{3} (x,t|a) &=& 2\int_{\max(0,1+x-a)}^x \rmd y \, \Big[ -2\partial_y  \mathbb P(1-y,t) \nn\\
& &\hspace{3.05cm} -2\partial_y\partial_t  \mathbb P(1-y,t) \nn\\
&& \hspace{3.05cm} +(1-y)\partial_y^2  \mathbb P(1-y,t)\Big] \nn \\
&=&\mathbb R(x,t) -\mathbb R\big(\!\max(0,1+x-a),t\big)  \ .
\eea
This integral is analogous to \eq{67}, with the difference that the lower boundary may be larger than 0; this domain of integration is restricted s.t.\ the process never touches the lower boundary. 
For  $a\ge 1+x$, this reproduces the probability $p_{1a}(x,t)$, 
\be
p_{1a}(x,t) = p_3(x,t|a)\Big|_{a\ge 1+x}\ .
\ee
Using our assumptions, $p_{3}  (x,t|a)$ can be simplified to 
\be
p_{3} 
 (x,t|a)
=\mathbb R(x,t) -\mathbb R ( 1+x-a,t\big)\ . 
\ee
To get to  the last line we used our assumption \eq{x-cond}. 

Similarly, the last term in Eq.~\eq{88} reproduces the function $p_{1b}(x,t)$ used above, when choosing
\be
p_{1b}(x,t) = p_{3}(a-x,t|a) \Big|_{a=1+x}  \ .
\ee
Note that \Eq{88} has manifestly the symmetry $x\to a-x$, both for $p_2$ and $p_3$.  Choosing  $a=1+x$, the sum of the latter terms becomes  $p_{1a}(x,t)+p_{1b}(x,t)$, making manifest the hidden symmetry between these terms. 

Finally, one checks that for $x$ satisfying condition \eq{x-cond}, $\int_0^\infty \rmd t\, P_{ T_1}^{\rm NN}(x,t)=1$, thus the probability \eq{88} is properly normalized. A numerical test is presented on Fig.~\ref{f:PT1NNweighted}.

\section{Conclusions and Open Problems}
\label{s:Conclusions and Open Problems} 
Let us come back to the image of a myopic foraging rabbit, and ask
when he is no longer hungry. Suppose there is a uniform food distribution. The rabbit starts with an empty stomach, does a random walk and eats everything he can get, until his stomach is full ($S=1$). The probability for this time   is the probability that the span reaches one for the first time, as given in \Eq{47}, and after.
But a real rabbit is burning food, so  add a (negative) drift, i.e. stop when  $S(t)-\mu t =1$. 
Curiously, this problem is much more difficult to solve, and we (currently) have  no analytical solution. One may be able to calculate the probability that the rabbit dies before having a full stomach, following the approach outlined in Ref.~\cite{BraySmith2007}. 

Another open problem is the generalization of the observables obtained here for correlated processes, as fractional Brownian motion. 
While the first moments of the   span distribution have been  obtained in an expansion \cite{Wiese2018} around $H=1/2$ (Brownian motion), the full distribution remains to be evaluated.

\vspace*{5mm}
\centerline{\bf ACKNOWLEDGMENTS}
\medskip
It is a pleasure to thank Olivier Benichou, Joachim Krug, Satya Majumdar, Tridib Sadhu and Sidney Redner for stimulating discussions. 

\vfill

\ifx\doi\undefined
\providecommand{\doi}[2]{\href{http://dx.doi.org/#1}{#2}}
\else
\renewcommand{\doi}[2]{\href{http://dx.doi.org/#1}{#2}}
\fi
\ifx\link\undefined
\providecommand{\link}[2]{\href{http://#1}{#2}}
\else
\renewcommand{\link}[2]{\href{http://#1}{#2}}
\fi
\providecommand{\arxiv}[1]{\href{http://arxiv.org/abs/#1}{#1}}

\tableofcontents

\end{document}